\documentclass[aps,10pt,a4paper,showpacs,pre,twocolumn]{revtex4-1}
\usepackage{amssymb}
\usepackage[dvips]{graphicx}
\usepackage{subfigure}
\usepackage[dvips]{color}
\newcommand{\mb}{\mathbf}
\newcommand{\mc}{\mathcal}

\begin{document}

\title{Using field theory to construct hybrid particle-continuum simulation schemes
with adaptive resolution for soft matter systems}
\author{Shuanhu Qi, Hans Behringer, Friederike Schmid}
\affiliation{Institut f\"{u}r Physik, Johannes Gutenberg-Universit\"{a}t Mainz, Staudingerweg 9, D-55099 Mainz, Germany}

\begin{abstract}
We develop a multiscale hybrid scheme for simulations of soft
condensed matter systems, which allows one to treat the system at
the particle level in selected regions of space, and at the
continuum level elsewhere. It is derived systematically from an
underlying particle-based model by field theoretic methods.
Particles in different representation regions can switch
representations on the fly, controlled by a spatially varying tuning
function. As a test case, the hybrid scheme is applied to simulate
colloid-polymer composites with high resolution regions close to the
colloids. The hybrid simulations are significantly faster than
reference simulations of a pure particle-based model, and the
results are in good agreement.
\end{abstract}

\maketitle

\section{Introduction}

Multiscale modeling is one of the central challenges in many areas
of materials science \cite{AA_CG,EEL07,cit1}. The properties of
modern materials are often determined by an interplay of structural
features and processes on length scales that span several orders of
magnitude. For example, many materials are heterogeneous on a nano-
or micrometer scale and filled with ``defects" -- internal
interfaces, droplets of a different phase, or nanoparticle fillers.
Theoretical descriptions must account for the microscopic structure
close to these defects as well as the larger scale structure of the
``bulk'' medium surrounding the defects \cite{cit2}.  To study such
systems, multiscale modeling approaches have been developed and
pursued for several decades, which employ a hierarchy of models to
describe the material properties at different coarse graining levels
\cite{Multiscale2}. One crucial issue in this context is
the coupling between models. The traditional approach has been to
couple them ``vertically", i.e., simulations of different models are
run independently and linked by parameter heritage. Nowadays,
``horizontal coupling" schemes are attracting growing interest,
where regions of different resolution coexist within one single
simulation system \cite{cit4}.  In particular, the adaptive
resolution models \cite{AdRes, Potential_control,
PIandCG,PinningPotential, HAdRes}, which allow free diffusion of
particles between regions of different resolution, are able to
dynamically couple information and to account for density
fluctuations and flow. The adaptive scheme is suitable for systems
with small regions requiring detailed investigation, while the
remaining large part only needs a computationally cheaper
coarse-grained description. Such systems are ubiquitous in soft
materials, e.g., chemical reaction systems, polymer solutions and
melts with interfaces, or composite materials.

On the microscopic side, materials are typically represented by particle-based
models (atomistic or coarse-grained). On the macroscopic side, continuum models
are commonly used (elastic models, phase field models, hydrodynamic models).
While horizontal coupling schemes have been developed both within the
``particle world" and the ``continuum world", linking the two still remains a
challenge. Hybrid particle-continuum schemes have been proposed where certain
molecules or components are treated permanently at the particle level, and
others permanently at the field level \cite{cit5, Sides, BD_DDFT}.  Other examples of
coupled schemes are ``Single Chain in Mean Field" simulation methods, where
particles move in the dynamically updated mean field of the surrounding
particles \cite{MS05b,SMK05,PF_MD}, or 'heterogeneous multiscale' schemes where
particle simulations are used to adjust the parameters of a continuum
simulation on the fly \cite{EEL07,cit7}. However, apart from proposals for
simple liquids \cite{FDC06,FDC07,DKP08}, the present authors are not aware of a
general scheme for complex fluids that would allow one to treat different
regions of space at either particle or continuum level in an adaptive
resolution sense.

With the present paper, we aim at closing this gap. We propose a
method to generate adaptive resolution schemes that link particle
and continuum representations of the same complex fluid in a
formally exact manner. Together with existing adaptive
particle-particle and continuum-continuum coupling schemes, our
method could potentially pave the way to integrated multiscale
treatments of complex fluids from the atomistic to the macroscopic
scale.
 
\section{Basic concept of the approach}

Our starting point on the particle side are models of the Edwards
type, which can be defined in terms of local densities. This
implies, in particular, that the interaction potentials are soft,
i.e., molecules can penetrate each other. Although the Edwards
models were originally introduced in the context of analytical
theory \cite{cit8}, they also proved to be efficient models for
computer simulations \cite{Offlattice,SMK05,Nearest,Hans}. The
partition function of an Edwards-type model can be rewritten exactly
as a fluctuating field theory \cite{cit9}, either by applying a
Hubbard-Stratonovich transformation (if the interactions are purely
quadratic in the densities), or, more generally, by inserting
unities (delta functions) in a Faddeev-Popov way
\cite{Schmid,Matsen,MS05,G_book}. Such fluctuating field models have
also been studied by computer simulations with considerable success
\cite{FTS1,FTS2,LKF08}. Even more importantly, fluctuating field
theories lend themselves to mean-field approximations, thus
providing a natural link between Edwards models and popular density
functional theories for complex fluids such as the Self-Consistent
Field (SCF) theory \cite{Schmid,G_book} or dynamic density
functional theories \cite{MS05,cit10,EPD}. These so-called
``molecular field'' theories are nonlocal continuum models, which
can be used directly for mesoscale simulations of complex fluids
\cite{HS06,HS08}, and which also provide an excellent starting point for
systematic derivations of simpler phase field theories \cite{cit11}.

Thus every particle-based Edwards model has a continuum model partner,
i.e., the corresponding molecular field model, which is equivalent
apart from a mean-field approximation. We note that the mean-field
approximation becomes accurate in dense systems, which is where the
transition from a particle-based to a field-based model is most attractive.
Moreover, the effect of fluctuations can often be included to some
extent even in a molecular field simulation \cite{FTS2,MS05}.

Our adaptive resolution scheme exploits this correspondence between
Edwards models and molecular field models . We will
construct a hybrid model that combines particle and field
representations of the same molecules, and a simulation scheme to
switch between representations depending on the position in space.
The switching probability is controlled by a spatially varying
virtual field $\Delta \mu({\bf r})$, which can be chosen at will. As
an example, we will study a polymer-colloid composite, with $\Delta
\mu({\bf r})$ chosen such that the particle representation dominates
close to the colloids, and the field representation far from the
colloids.

We will now describe the basic idea of our approach. Technical
details are given in the Appendix A. For simplicity, we
consider a one-component system of $n$ polymers (labelled $\alpha$)
with $N$ monomers (labelled $j$). Our starting point is the
canonical partition function
  \begin{equation}
    \label{eq:z_particle}
    {\cal Z}=\frac{1}{n!} \int \prod_{\{\alpha j\}} {\rm d} {\bf R}_{\alpha, j}
    \exp\{-\mc H_0-\mc H_\mathrm{nb}\},
  \end{equation}
where the integrals run over all monomer positions $\mb R_{\alpha
j}$, $\mc H_0$ denotes the Gaussian spring energy of the chains, and
$\mc H_\mathrm{nb}$ describes the non-bonded interactions in terms
of an Edwards Hamiltonian. Here and throughout, the energy unit is
chosen $1/k_\mathrm{B}T\equiv 1$. Based on the partition function
(\ref{eq:z_particle}), the hybrid particle-field model is now
constructed in three steps.

In the first step, the polymer chains are partitioned into two
different (virtual) species, which we name p-chains and f-chains.
This is done by attaching an additional virtual variable
$\tau_{\alpha j}\in\{0,1\}$ to each monomer.  A chain $\alpha$ is
said to be an f-chain if $\sum_j \tau_{\alpha j}=0$, otherwise it is
called a p-chain. The virtual variables $\tau_{\alpha j}$ are
introduced by inserting the exact identity
\begin{equation}
\label{eq:identity}
\sum_{\tau_{\alpha j}=0}^1 \exp\left[ \tau_{\alpha j}
\Delta \mu(\mb r) - \ln\left(\mathrm{e}^{\Delta\mu (\mb r)} 
+ 1\right)\right] = 1
\end{equation}
in the partition function, Eq.\ (\ref{eq:z_particle}). This 
couples them to the virtual field $\Delta \mu({\bf r})$,
and the latter can be used to control the fraction of f- and p- chains 
at a given position $\mb r$.  We note that we are free to choose
the field $\Delta \mu({\bf r})$ as we like, since the identity,
Eqn. (\ref{eq:identity}), is exact.

The second step is to treat the p-chains and the f-chains by
different representations. We keep the particle description for the
p-chains, but convert the description of f-chains into a field
representation. This is done in the usual Faddeev Popov way by
inserting appropriate identity operators (see Refs.\
\cite{Schmid,Matsen,MS05} or Appendix A). As a result,
the particle degrees of freedom of the f-chains are replaced by
fluctuating fields $\phi_\mathrm{f}$ and $\omega_\mathrm{f}$.

The resulting expression for the partition function is formally
equivalent to Eq.\ (\ref{eq:z_particle}), but it cannot be sampled
efficiently. Therefore, the third step is to introduce convenient
approximations that speed up the numerical calculations. Here we use
a saddle point evaluation \cite{Schmid,EPD,GaussianFluc} of the
$\omega_\mathrm{f}$ integral, and just keep the $\phi_\mathrm{f}$
fields. Such a mean-field type treatment only influences the
contributions from the f-chains. It amounts to a kind of
`coarse-graining' in the low-resolution region. If the density of
the medium there is high, the mean-field approximation is known to
describe the system very well. The physics in which one is
interested, however, is extracted from the high-resolution part
where the polymers are still represented by particles, for which no
approximations were used.

\section{Application example: Polymer-colloid-nanocomposite}

As an application of our hybrid model, we study a complex composite
system containing two nanocolloids that are coated uniformly with
homo-brush polymers and immersed in a melt of $n_t$ A-B diblock
copolymers. Each free polymer consists of $N=20$ monomer beads, with
$N_A=10$ A-beads and $N_B=10$ B-beads, and each brush polymer
contains 10 monomer beads. One colloid is coated with
A-homopolymers,  the other with B-homopolymers. The non-bonded
Edwards Hamiltonian for this system is given by
\begin{equation}
\frac{\mc H_{\text{\scriptsize nb}}V}{n_t}=\chi N\int d\mb r
\hat\phi_A\hat\phi_B
+ \kappa N\int d\mb r\big[\phi_0-\hat\phi_A-\hat\phi_B\big]^2,
\end{equation}
where the Flory-Huggins parameter $\chi N=9$ measures the
incompatibility of monomers $A$ and $B$, $\kappa N=10$ is the
compressibility. The configuration dependent densities of monomers
$A$ and $B$ are denoted by $\hat{\phi}_A$ and $\hat{\phi}_B$, and
$\phi_0$ is the reference monomer density in the bulk fluid.
Furthermore, monomers are not allowed to enter the colloids. All
lengths are measured in the units of the mean radius of gyration of
free (ideal) polymers $R_g\equiv\sqrt{Nb^2/6}$.

We consider a system consisting of $n_t=20000$ free polymers in a
simulation box of size $L_x=L_y=8$, $L_z=32$, resulting in an
invariant degree of polymerization \cite{MS05} $\sqrt{\bar\mc
N}=\frac{\bar\rho R_e^3}{N}\simeq 144$ (here $\bar\rho$ is the
average bead density of the free polymers and $R_e$ is the mean
end-to-end distance of free polymers). The system is discretized in
cubic cells of side length $0.25$, which are used both for the
field-theoretic calculations and the evaluation of local monomer
densities. Two colloids of radius $R_g$ are placed on the centerline
$x=y=0$ at fixed distance from each other. They are coated with
$n_b$ graft polymers with either $n_b=37$ (low grafting density) or
$n_b=143$ (high graft density). The densities of p-chains are
calculated using the particle-to-mesh method \cite{Nearest}.

\begin{figure}[t]
  \centering
  \subfigure[]{
    \label{fig:subfig:a} 
    \includegraphics[angle=0, width=6.0cm]{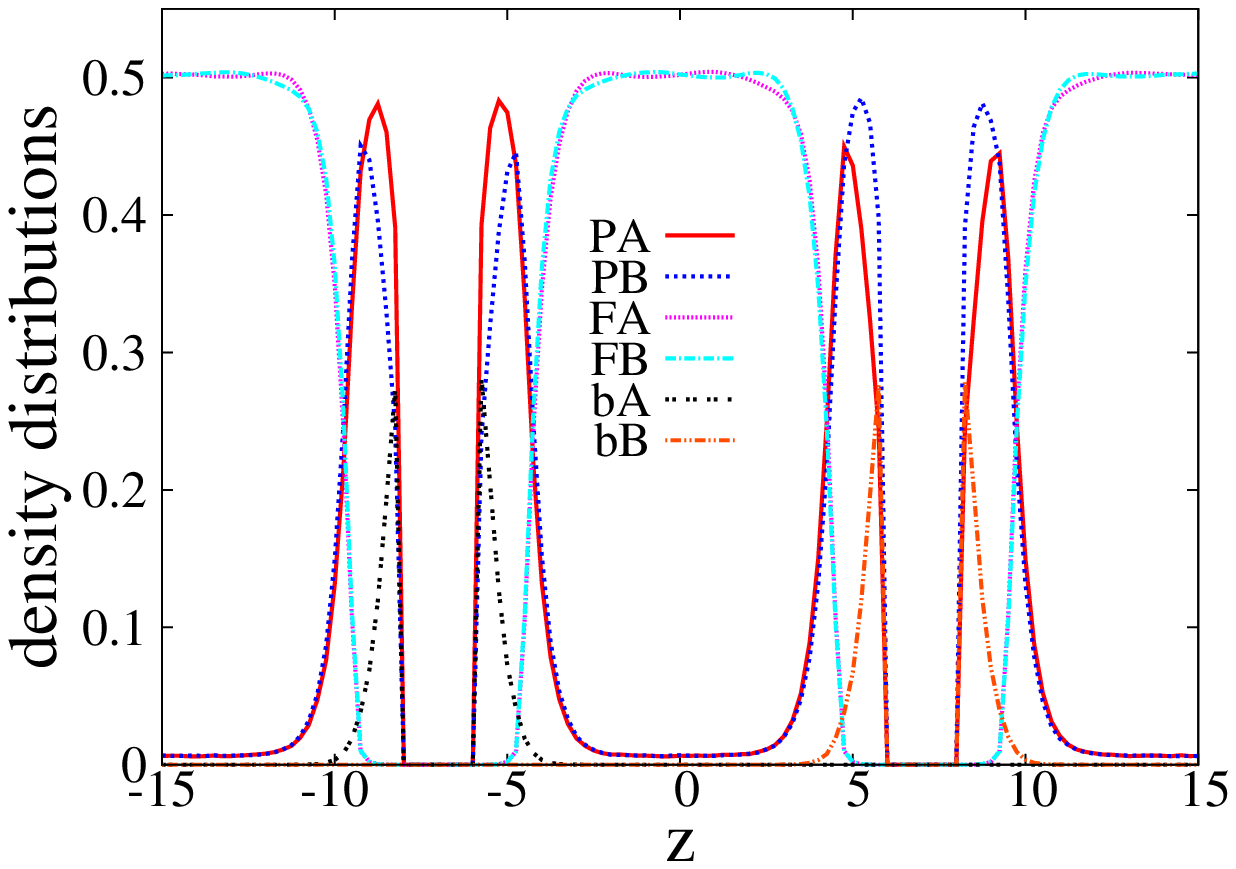}}
  \subfigure[]{
    \label{fig:subfig:b} 
    \includegraphics[angle=0, width=6.cm]{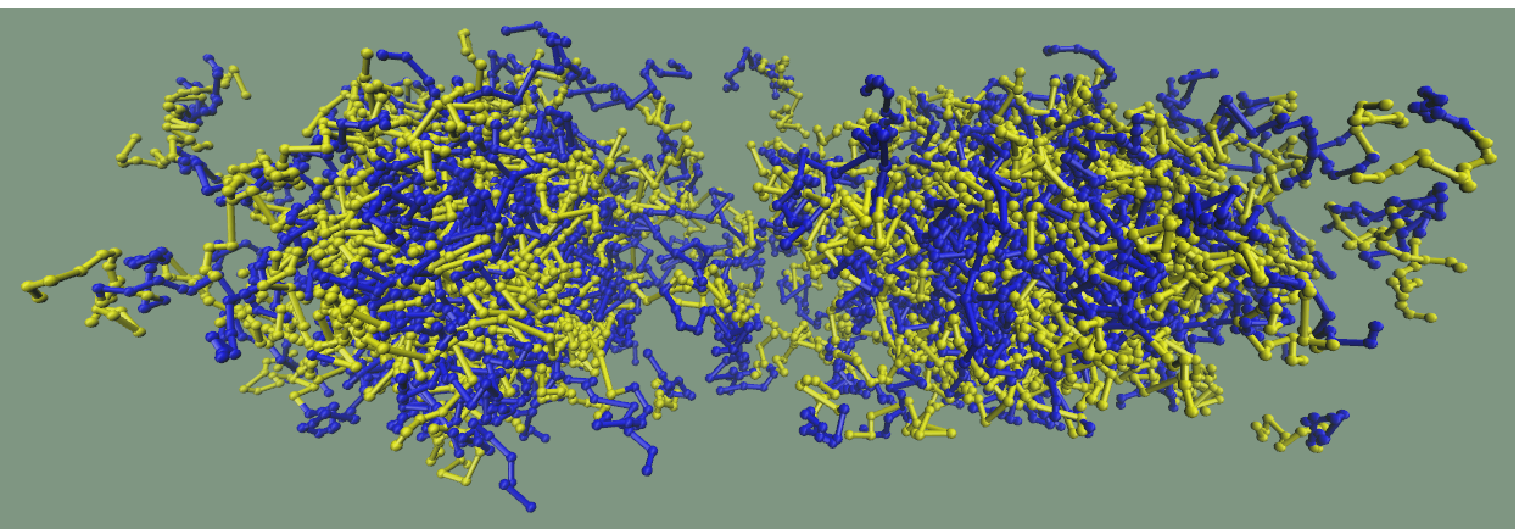}}
  \caption{Density profiles of free p-chains (PA, PB), f-chains (FA, FB), and
brush polymers (bA, bB) along the line
$x=0,y=0$ with one colloid located at (0,0,-7), and the other at
(0,0,7) (a). Corresponding snapshot showing just p-polymers (b). The grafting density is
$n_b=143$. The average number of p-chains is about 2100 (out of $n_t
= 20000$ free polymers total). }
  \label{fig:densityProfileSnapshot} 
\end{figure}

To determine a suitable tuning function $\Delta \mu({\bf r})$, we
must first choose a pair of values $\Delta \mu_f$ and $\Delta
\mu_p$, for which a homogeneous bulk system is occupied almost
exclusively by f-chains (fields) or p-chains (particles),
respectively. A good choice in our system is $\Delta \mu_f=-4$ and
$\Delta \mu_p = 1.2$. The function $\Delta \mu({\bf r})$ then
interpolates between $\Delta \mu_p$ close to the colloids and
$\Delta \mu_f$ far from the colloids. Specifically, we used a step
profile, $\Delta\mu (r)=\Delta \mu_p+(\Delta \mu_f - \Delta \mu_p)
\Theta(r-r_c)$ with the Heaviside step function $\Theta$, where $r$
is the distance to the closest colloid and the shell thickness was
chosen $r_c = 2.5 R_g$. Figure \ref{fig:densityProfileSnapshot}
shows the density profiles of p- and f-chains along the line $x=0$,
$y=0$, along with a snapshot of the particle chains in the system.
One can see that particle chains dominate close to the colloid,
while in the bulk region far from the colloid, the polymers are
mostly represented by fields. The total volume of the particle
region is roughly $\sim 130 R_g^3$. 

The system was studied using a Monte Carlo simulation method which
includes three types of updating steps: (I) update the particle
configurations, (II) update the fields using a dynamic density
functional scheme, (III) update the $\{\tau\}$ configurations and
switch the chain identities accordingly. Moves (I) and (III) are
accepted according to the appropriate Metropolis criterion
\cite{MC_book}. To assess the performance of the hybrid model, we
have also carried out reference simulations of the same system in
pure particle representation. The hybrid simulations were roughly
three times faster than the simulations of the particle model.

\begin{figure}[t]
  \centering
  \subfigure[]{
    \includegraphics[angle=0, width=4.3cm]{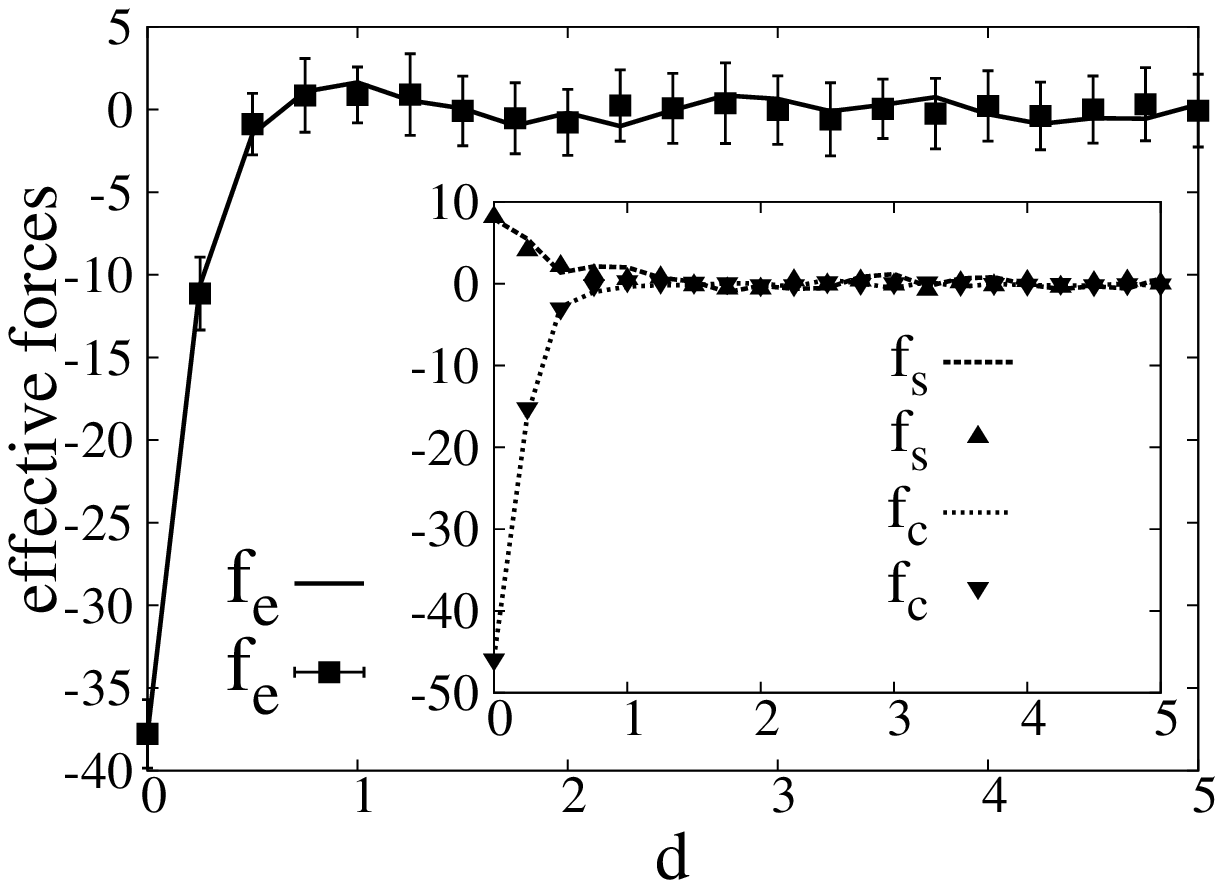}}
  \hspace{-0.5cm}
  \subfigure[]{
    \includegraphics[angle=0, width=4.3cm]{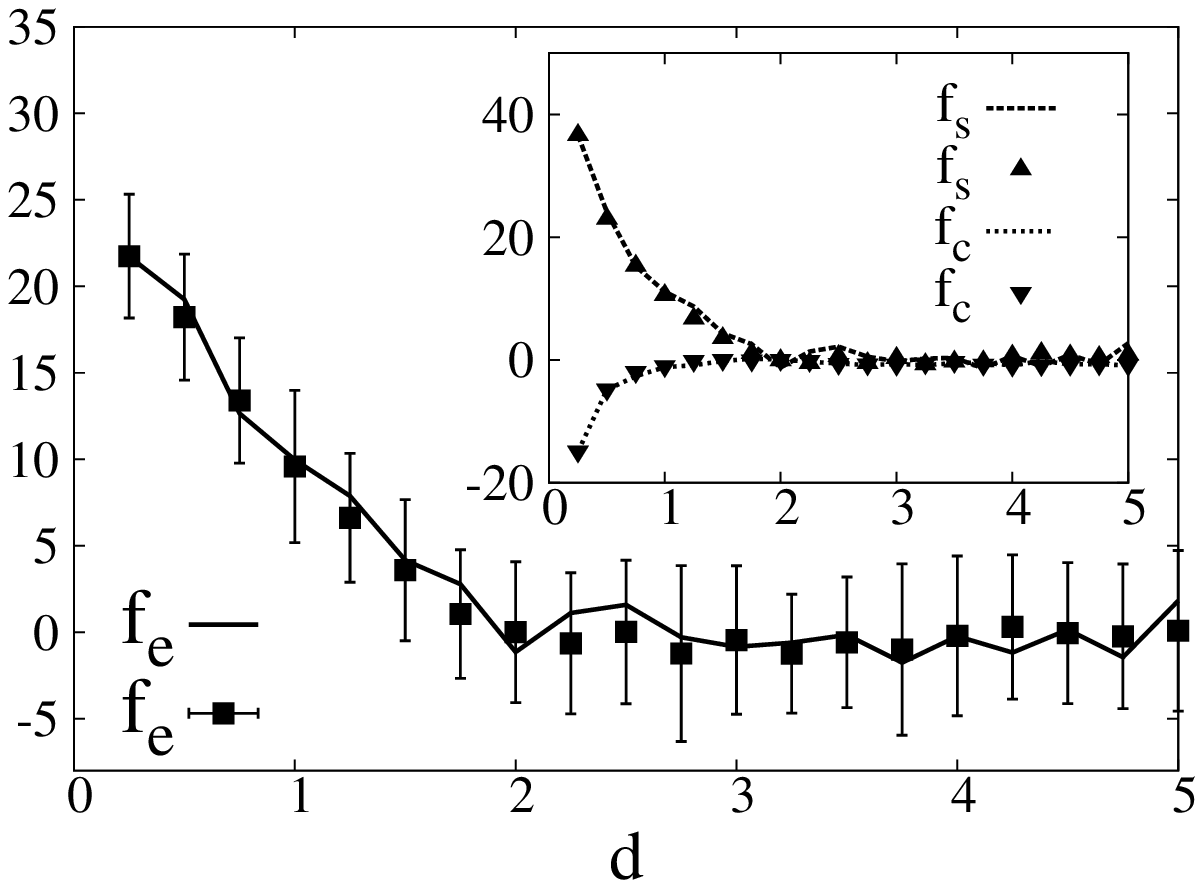}}
  \caption{Effective force $f_e$ between colloids, given by the sum of spring
    force $f_s$ (inset) and contact force $f_c$ (inset) in unit of
    $\frac{k_BT}{R_g}$ as a function of the distance between the two
    colloids $d$ with $n_b=37$ (a) and $n_b=143$ (b), calculated
    with the hybrid model (lines), and the corresponding
    pure particle-based model (symbols). The error bars for the hybrid
    model are comparable to those for the particle model.}
  \label{fig:effectiveForce} 
\end{figure}

We first consider the effective force between colloid particles
\cite{Cinteraction1,C_DR, C_DR2}, which determines the stability and
uniformity \cite{Cinteraction2} of the composite material. It is
given by the mean total force acting on one colloid if the other one
is kept fixed at a certain distance, and it has two contributions:
The mean spring force from the graft polymers, and the mean contact
force due to the unsymmetrical collisions of the beads around the
colloid. The latter can be expressed \cite{CForce1,CForce2} as an
integral over the surface $A$ of the colloid $\vec{f}_c=- \int {\rm
d}^2A \: \vec{n} \: \rho(A)$, where $\vec{n}$ is the surface normal,
and $\rho(A)$ the local density of beads at the surface, which
includes p-, f-, and graft chains. Figure \ref{fig:effectiveForce}
shows the effective total force as a function of the distance $d$
between the two colloids for different numbers of brush polymers. At
low grafting density, the colloids attract each other due to the
depletion effect. At high grafting density, the brushes induce an
entropic repulsion. This is the regime where the brush stabilizes
the colloidal system. For comparison, we also show the results for
the reference pure particle system. They are in good agreement with
the results from the hybrid model.

\begin{figure}[t]
\centerline{\includegraphics[angle=0,scale=0.5,draft=false]{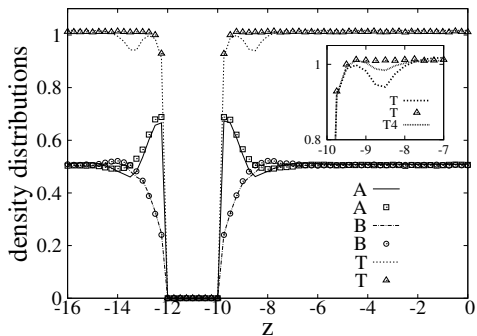}}
\caption{\label{fig:densityProfile} Density distributions of all A-beads (A),
  B-beads (B) and total (T) obtained from the hybrid model (lines) and
  from the pure particle model (symbols) at the line $x=0$, $y=0$ for two
  colloids with $n_b=143$ graft polymers. The inset shows a blowup of the
  total density profile in the dip region, with results for higher polymer
  density $n_t=40000$ for comparison (T4).}
\end{figure}

Next we investigate how the colloids  perturb the surrounding
polymer medium. Since the $\chi$
parameter ($\chi N= 9$) is below the order-disorder transition (ODT)
point ($(\chi N)_{ODT} \gtrsim 10.5$ \cite{Leibler,ODT_muller}), the polymer
melt is homogeneous in the bulk. Close to the colloid surface, we
observe colloid induced ordering. Figure \ref{fig:densityProfile}
shows the density profiles for all A-beads, all B-beads and the
total density along the line $x=0$, $y=0$ for a systems containing
two at positions (0,0,-11) and (0,0,11), respectively. Only the
density profile in half the system is shown, since the other half is
symmetric.

The results obtained from the pure particle model, also
shown in Fig. \ref{fig:densityProfile}, are again in good agreement,
except for a small density dip in the p-f interfacial region.
A similar density dip, with comparable magnitude, has also been
found in other adaptive resolution schemes
\cite{PIandCG,PinningPotential}. In our case, it can be related to
the mean-field approximation: When increasing the density of free
polymers, the dip becomes smaller (see inset in
Fig.~\ref{fig:densityProfile}). It can be reduced by making the 
``interfacial region" between p- and f-regions broader,
e.g., choosing a smooth tanh-like profile for $\Delta \mu(\mb r)$
instead of the simple step function used here. A detailed analysis
of these effects will be published elsewhere.
Another possibility is to follow Ref.\ \cite{PinningPotential} and
introduce an additional potential in the interfacial region.

By using a sharply varying tuning function that produces a
relatively pronounced density dip, we can assess its influence on
the other structural properties of interest. Despite the artifact,
colloidal forces are still reproduced accurately by the hybrid
model, and the relative distribution of A and B monomers around the
colloid is in good agreement with that in the pure particle model.
Thus the presence of the artifact seems acceptable in the present
system. It might cause problems if one adds small molecules, which
might accumulate at the p-f ``interface" and whose transport
properties across the interface might be altered. In such
simulations, the artifact should be removed, e.g., by choosing a
tuning function that varies sufficiently slowly.

\section{Summary}

In summary, we have developed a hybrid particle-field scheme for
simulations with adaptive resolution, which dynamically couples
finer particle degrees of freedom with coarser field degrees of
freedom. The scheme has been tested at the example of a
nanocolloid-polymer composite and verified by comparing results from
hybrid simulations to results from pure particle simulations.  The
new scheme has been derived using a field-theoretic methodology that
can be applied very generally to molecular systems without hard core
interactions. Hence the approach should be widely applicable for all
materials which can be described by Hamiltonians with soft
interactions, i.e., typically soft matter systems.

In the present application, the hybrid simulations were found to be
roughly three times faster than the corresponding pure particle
simulations. The speedup will be even bigger in large systems
containing only small regions where a particle representation is
necessary. Field-based simulations have the advantage that the
computational costs do not increase with the number of molecules.
The hybrid approach will thus be particularly attractive for
simulations of dense systems, or of polymers with large
polymerization index, where particle simulations become expensive
compared to field-based simulations. Compared to pure field-based
simulations, the hybrid simulation method has the advantage that
inclusions and surfaces can be modeled accurately without having to
resort to approximate effective descriptions \cite{Sides}.

Since we have focused on equilibrium static properties in this work,
we have used a Monte Carlo simulation method to sample the partition
function. More realistic dynamical models can be implemented as
well. For example, overdamped Brownian particle simulations can be
combined in a straightforward manner with a dynamic density
functional that reproduces Rouse dynamics in field-based simulations
\cite{EPD}. This model would however neglect hydrodynamic
interactions. In order to include these, one could combine a
molecular dynamics scheme for the particles \cite{PF_MD} with a
momentum-conserving field-based simulation scheme \cite{Hall,Zhang}.
Such an approach would allow one to use the hybrid model for
studying dynamics and flow phenomena in complex fluids.

Another promising direction for future developments will be to
replace the tuning function $\Delta \mu({\bf r})$ that controls the
local particle and field content by a function that depends on local
densities or order parameters, $\Delta \mu(\rho({\bf r}))$. The high
resolution regimes can then adjust on the fly to the local
configurations.

\begin{acknowledgments}
We thank S. A. Egorov, S. Meinhardt, S. Dolezel, L. Zhang, and J.
Zhou for helpful discussions and suggestions. This work was funded
in part by the German Science Foundation. The simulations were run
on the computer cluster Mogon at the University of Mainz.
\end{acknowledgments}

\begin{appendix}

\section{Construction of the hybrid particle-continuum scheme: Technical details}\label{appendixA}

For simplicity, we derive the hybrid scheme for a simple polymeric
system of $n$ Gaussian chains (labeled $\alpha$) of one (chemical)
type with $N$ monomers (labeled $j$) in a volume $V$ at temperature
$T$. In the following, energies are given in units of
$1/k_\mathrm{B} T$, and lengths in units of the radius of gyration
of ideal chains, $R_g=\sqrt{{Nb^2}/{6}}$, where $b$ is the
statistical segment length. The total energy $\mc H$ includes the
Gaussian spring energy
  \begin{equation}
   {\mc H_0} = \sum_{\alpha=1}^n \frac{N}{4}\sum\limits_{j=1}^{N-1}(\mb
   R_{\alpha j}-\mb R_{\alpha {j+1}})^2,
  \end{equation}
where $\mb R_{\alpha j}$ denotes the position of monomer $j$ in
chain $\alpha$, and non-bonded contributions described by an Edwards
term that is defined in terms of local densities, e.g.,
  \begin{equation}
    {\mc H_{\text{\scriptsize nb}}}=\frac{nv}{2V}\int d\mb r\hat\phi^2
  \end{equation}
with excluded volume parameter $v>0$. Here $\rho_0 \hat \phi =
\sum_{\alpha,j} \delta (\mb r - \mb R_{\alpha,j})$ is the
configuration dependent density, which includes contributions from
all monomers $j$ of chains $\alpha$ at positions $\mb R_{\alpha,j}$,
and $\rho_0 = nN/V$ is the mean density. The total partition
function is then given by
  \begin{equation}
   \label{eq:z_1}
   \mc Z= \frac{1}{n!} \int\prod_{\alpha,j} {\rm d}
   \mb R_{\alpha,j}\mathrm{e}^{-\mc H_0-\mc H_\mathrm{nb}}.
  \end{equation}

In the first step, we partition all chains into two different
species, named p-chains and f-chains. This is done by attaching an
additional variable (label) $\tau_{\alpha j} \in\{0,1\}$ to each
monomer. This spin like variable $\tau$ can be coupled to the tuning
function $\Delta\mu(\mb r)$ by exploiting the identity
 \begin{equation}
    \sum_{\tau=0}^1 \exp\Big[\tau\Delta\mu(\mb r)
      -\ln\big(\mathrm e^{\Delta\mu(\mb r)}+1\big)\Big]=1.
 \end{equation}
This identity holds for any form of $\Delta\mu(\mb r)$ at any
position $\mb r$, so our method is not restricted to some specific
forms of $\Delta\mu(\mb r)$. Inserting this identity for each
$\tau_{\alpha,j}$ into the partition function, Eq.\ (\ref{eq:z_1}),
one gets
 \begin{equation}
  \mc Z = \frac{1}{n!}\sum_{\{\tau_{\alpha,j}\}} \int \prod_{\{\alpha,j\}}
  {\rm d} \mb R_{\alpha,j} \,\mathrm{e}^{-\mc H_{\Delta \mu}},
 \end{equation}
with
\begin{equation}
\mc H_{\Delta\mu} = \mc H_0 + \mc H_\mathrm{nb} +
\sum\limits_{\{\alpha,j\}}
   \left[U_{\Delta \mu}(\bf R_{\alpha,j})
    - \tau_{\alpha,j}\Delta\mu(\mb R_{\alpha,j})
    \right]
\end{equation}
 where we have defined $U_{\Delta \mu}({\bf r}) :=
\ln(\mathrm{e}^{\Delta\mu(\mb r)}+1)$.  This partition function
describes a system with additional auxiliary degrees of freedom
$\tau_{\alpha,j}$, which however have no physical meaning. The
construction ensures that the physics is not changed, compared to
the original system.

Let us now assume that we have a given partitioning of molecules
into two virtual identities, namely $n_\mathrm{p}$ p-chains and
$n_\mathrm{f}$ f-chains (with $n = n_\mathrm{p} + n_\mathrm{f}$).
Then the non-bonded energy is given by
 \begin{equation}
 \mc H_{\mathrm{nb}}= \frac{nv}{2V} \int d \mb r (\hat \phi_\mathrm{p}^2
                       + 2 \hat \phi_\mathrm{p}\hat\phi_\mathrm{f}
                       + \hat\phi_\mathrm{f}^2),
 \end{equation}
with $\hat\phi_{\mathrm{p}}$  the configuration dependent monomer
density for the p-chains, and $\hat\phi_\mathrm{f}$ analogously for
the f-chains. Obviously, the value of $\mc H_\mathrm{nb}$ for a
given system configuration does not depend on the partitioning into
p- and f-chains.  Therefore, one can use a different partitioning
for each set of $\{\tau_{\alpha,j}\}$.  We shall use the rule that a
chain $\alpha$ is an f-chain if $\tau_{\alpha,j} = 0$ for all
monomers $j$, otherwise it is a p-chain.  The function $\Delta
\mu({\bf r})$ then tunes the statistical weight of a particular p-
and f-chain partitioning in the partition function. The p-chains and
f-chains act like two different species, thus the system becomes
semi-grand-canonical.

The procedure so far has only complicated the notation. However, the
p-chains and the f-chains can now be treated by different
representations.  We keep the particle description for the p-chains,
but convert the description of f-chains into a field representation.
This is done technically in the usual way by inserting an identity
operator \cite{Schmid}
 \begin{equation}
  \openone \propto \int D \phi_\mathrm{f} \int
  D\omega_\mathrm{f} \exp\left(\frac{n}{V}\int d\mb r \,i\omega_\mathrm{f}(\mb
r)[\phi_\mathrm{f}(\mb r) - \hat\phi_\mathrm{f}(\mb r)]\right),
 \end{equation}
for the local densities $\hat\phi_\mathrm{f}$ of the f-chains, where
$\phi_\mathrm{f}$ now denotes the associated density field and
$i\omega_{\mathrm{f}}$ is the conjugate field. The particle degrees
of freedom of the f-chains can be integrated out resulting in a
single chain partition function $Q_f[i\omega_\mathrm{f},\Delta\mu]$
 \begin{widetext}
 \begin{equation}
 \label{eq:q}
 \mc Q_f= \mc N \: {\int \prod_j {\rm d}\mb R_j \exp\left\{
   -\frac{N}{4}\sum\limits_{j=1}^{N-1}(\mb R_j-\mb R_{j+1})^2
   -\frac{1}{N}\sum\limits_{j=1}^{N}i\omega_f[\mb R_j]
   -\sum\limits_{j=1}^{N}\ln[e^{\Delta\mu(\mb R_j)}+1]\right\}},
 \end{equation}
 \end{widetext}
and their associated physics is described by the fields
$\phi_\mathrm{f}$ and $\omega_\mathrm{f}$. In Eq.\ (\ref{eq:q}), a
normalization factor
 \begin{equation}
 \mc N^{-1} = \int \prod_j {\rm d}\mb R_j \exp\left\{
    -\frac{N}{4}\sum\limits_{j=1}^{N-1}(\mb R_j-\mb R_{j+1})^2\right\}
 \end{equation}
has been included for numerical convenience. The partition function
then can be written in the form
 \begin{equation}
  \label{eq:partitionFunction}
  \mc Z = \sum_{\{\tau_{\alpha,j}\}} \int
  D \phi_\mathrm{f} \int D\omega_\mathrm{f} \int
  \prod_{\{\alpha_\mathrm{p},j\}}
  \mb d\mb R_{\alpha_\mathrm{p},j} \,\mathrm{e}^{-\mc H_\mathrm{eff}},
 \end{equation}
where the index $\alpha_\mathrm p$ indicates that the
configurational integral $\int \prod_{\{\alpha_\mathrm{p},j\}} {\rm
d} \mb R_{\alpha_p j}$ now runs over the monomers of p-chains only.
The effective Hamiltonian $\mc H_\mathrm{eff}$ for a given
configuration $\{\tau_{\alpha_\mathrm{p},j}\}$ with particle
positions $\{\mb R_{\alpha_\mathrm{p},j}\}$ and field values
$\phi_\mathrm{f}$ and $\omega_\mathrm{f}$ can be split into three
contributions
 \begin{equation}
  \mc H_\mathrm{eff} = H_\mathrm{p} + H_\mathrm{f} + H_\mathrm{pf}.
 \end{equation}
Here $H_\mathrm{p}$ corresponds to the pure contributions of
p-chains, including in particular their interaction with the virtual
potentials $\Delta\mu ({\bf r})$ and $U_{\Delta \mu}({\bf r})$,
  \begin{widetext}
  \begin{equation}
    H_p=-\sum_{\{\alpha_p,j\}}\Delta\mu(\mb R_{\alpha_p,j})\tau_{\alpha_p,j}
     +\sum_{\{\alpha_p,j\}}\ln U_{\Delta\mu}(\mb R_{\alpha_p,j})
      +\frac{N}{4}\sum_{\{\alpha_p,j\}}\Big[\mb R_{\alpha_p,j}-\mb R_{\alpha_p,j+1}\Big]^2
      +\frac{nv}{2V}\int d\mb r\hat\phi_p^2(\mb r),
  \end{equation}
  \end{widetext}
$H_\mathrm{f}$ describes the pure contribution of f-chains in field
representation,
  \begin{equation}\label{H_f}
    H_\mathrm{f} = \frac{nv}{2V} \int d \mb r
    \phi_\mathrm{f}^2 - \frac{n}{V} \int d \mb r \,i
    \omega_\mathrm{f}\phi_\mathrm{f} - n_\mathrm{f}\ln
   Q_\mathrm{f}[i\omega_\mathrm{f},\Delta\mu],
  \end{equation}
and finally, the coupling term is given by
  \begin{equation}
   H_\mathrm{pf} = \frac{nv}{V} \int d \mb r \hat \phi_\mathrm{p}\phi_\mathrm{f}.
  \end{equation}
The partition function given by Eq.~(\ref{eq:partitionFunction}) is
our final, and formally exact expression of the partition function
for the present hybrid particle-continuum scheme. This partition
function contains both the particle and continuous field degrees of
freedom.

Unfortunately, the partition (\ref{eq:partitionFunction}) cannot be
sampled efficiently due to the imaginary contribution of $i
\omega_f$, which creates a sign problem (an oscillating integrand).
This problem is well-known in field-theoretic polymer simulations
\cite{FTS1}. It can be overcome by using the (computationally
expensive) Complex Langevin (CL) simulation method \cite{cit9}, but
this comes at the expense of having to introduce complex density
fields. Hence combining the CL method with particle simulations is
not straightforward.

However, most field-based simulation methods operate with real
density fields, which is made possible by employing additional
(mean-field) approximations. For example, in binary polymer blends,
the main effect of fluctuations was found to be sampled correctly by
an approach which treats the integral over $\omega_\mathrm f$ fields by a
saddle point integral and just samples the densities $\phi_\mathrm f$
\cite{FTS2}. This can be done within a suitable dynamic density
functional scheme \cite{MS05}. In the present work, we go one step
further and also neglect the fluctuations of $\phi_\mathrm f$ by setting the
noise in the dynamic density functional equations to zero. Such a
treatment is known to become accurate in the limit of high polymer
densities, or high invariant degree of polymerization \cite{FTS2}.
Our approach should be efficient in simulations where large parts of
the simulation volume can be treated safely at the (dynamic) mean field
level. For example, in phase separated polymer solutions, regions with high
polymer densities can be treated at the field level, and regions with low
densities at the particle level.

\section{Simulation method used in this work}

Our polymer/colloid composite was sampled using a Monte Carlo
method. The algorithm includes three different updating steps.
\begin{enumerate}
 \item For a given configuration of virtual spins
 $\{\tau_{\alpha j}\}$ and associated partitioning into
 p- and f-chains, and for given field degrees of freedom
 $\phi_\mathrm f$, the particle (monomer) positions $\mb R_{\alpha j}$
 are updated by local moves, which are accepted or rejected
 according to a Metropolis criterion.

 \item For the same $\{\tau_{\alpha j}\}$ the field degrees of
 freedom are updated while the particle conformations are kept
 fixed. Here we use a local relaxation scheme with noise set to zero,
 which amounts to a mean-field approximation where thermal
 fluctuations of the $\phi_\mathrm f$ are neglected. Specifically,
 our updating scheme is a variant of external potential
 dynamics \cite{EPD}: The densities are parameterized by the
 conjugate saddle fields $\omega_f$, which are updated according
 to $\omega_{\mu\mathrm  f} \to \omega_{\mu\mathrm f} + {\rm d}t \:
 \frac{\delta \mc H_\mathrm{eff}}{\delta \phi_{\mu\mathrm f}}$,
 where $\mu=A,B$, and ${\rm d}t$ is a parameter controlling
 the step length. This update also involves an evaluation of the
 propagator in Eq.(\ref{H_f}), which is done with a pseudo-spectral
 method \cite{G_book}.

  \item The configuration of the auxiliary variables
  $\tau_{\alpha j}$ is updated. This implies that p-chains may turn
  into f-chains and vice versa. In those cases p-chains are
  physically removed or inserted into the system, and the number
  $n_\mathrm f$ of f-chains changes accordingly. In case of a p-chain switching to an
  f-chain, the p-chain will be removed from the system, while in case of an f-chain
  switching to a p-chain, the new generated p-chain will be added to the system. In each
  possible switch, we only deal with one p-chain and one f-chain.
  In the present work, new chains were generated randomly with Gaussian distributed bonds.
  More sophisticated schemes such as configurational bias Monte
  Carlo moves \cite{MC_book} are conceivable as well. Trial moves
  are accepted or rejected according to a Metropolis criterion. Note that the field
  $\omega_\mathrm{f}$ and thus the propagator $Q_\mathrm{f}$ remain fixed in this step.

\end{enumerate}
In our simulations, one ``Monte Carlo step" included on average one
trial move of $\mb R_{\alpha j}$ per (particle) monomer ($(\alpha,
j)$, 2000 trial switches of a variable $\tau_{\alpha j}$
(corresponding to one attempted p-f switch per ten chains in our
system of $20,000$ chains), and the fields were updated every third
Monte Carlo step.

\end{appendix}

\end{document}